\documentclass[11pt]{article}
\pdfoutput=1
\usepackage{jheppub}
\usepackage{tabu}
\usepackage[vcentermath]{youngtab}
\usepackage[usenames,dvipsnames,table]{xcolor}
\usepackage{graphicx,amsmath,amssymb,amsthm,multirow,array,bm,bbm,esint}
\usepackage[mathscr]{eucal}
\usepackage{epsf,amsfonts}
\usepackage{slashed}
\usepackage{ulem}
\usepackage[numbers,sort&compress]{natbib}
\usepackage{minibox}
\usepackage{rotating}
\usepackage{pdflscape}
\usepackage{array,tikz-cd}
\usetikzlibrary{decorations.pathmorphing}
\usetikzlibrary{decorations.pathreplacing}

\usepackage{subfig}
\usepackage{float}
\definecolor{rust}{rgb}{0.8,0.2,0.2}

\usepackage{soul,xcolor}
\setstcolor{red}

\newcommand{\RR}{\mathbb{R}}
\newcommand{\CC}{\mathbb{C}}
\newcommand{\SSS}{\mathrm{S}}
\newcommand{\TT}{\mathrm{T}}
\newcommand{\dd}{\mathrm{d}}
\newcommand{\ii}{\mathrm{i}}
\newcommand{\erm}{\mathrm{e}}
\renewcommand{\Re}{\operatorname{Re}}
\renewcommand{\Im}{\operatorname{Im}}

\DeclareMathOperator{\Res}{Res}
\DeclareMathOperator{\csch}{csch}

\newcommand{\figref}[1]{Fig.~(\ref{fig:#1})}

\newcommand{\secref}[1]{Sec.~\ref{sec:#1}}
\renewcommand{\eqref}[1]{(\ref{eq:#1})}

\title{Chaos and pole skipping in CFT$_2$}

\author{David M.~Ramirez}
\affiliation[]{Brown Theoretical Physics Center and Department of Physics,  \\
Brown University, Providence, RI 02912-1843 USA}

\emailAdd{david\_ramirez@brown.edu}

\abstract{Recent work has suggested an intriguing relation between
  quantum chaos and energy density correlations, known as pole
  skipping. We investigate this relationship in two dimensional
  conformal field theories on a finite size spatial circle by studying
  the thermal energy density retarded two-point function on a
  torus. We find that the location $\omega_* = \ii \lambda$ of pole
  skipping in the complex frequency plane is determined by the central
  charge and the stress energy one-point function
  $\langle T\rangle$ on the torus. In addition, we find a
  bound on $\lambda$ in $c>1$ compact, unitary CFT$_2$s identical to
  the chaos bound, $\lambda \leq 2\pi T$. This bound is saturated in
  large $c$ CFT$_2$s with a sparse light spectrum, as quantified by
  \cite{Hartman:2014oaa}, for all temperatures above the dual
  Hawking-Page transition temperature.}

\begin{document}

\maketitle


\section{Introduction}
\label{sec:intro}

Recent years have seen tremendous progress in understanding the
chaotic properties of many body quantum systems. While long believed
to play a role in microscopic mechanisms for fundamental properties
such as transport and thermalization, chaos has traditionally been
difficult to quantify in general many body systems. Exciting
developments following the study of chaos in gravitational contexts
have illustrated the role of out-of-time-ordered correlation functions
(OTOCs) in diagnosing chaotic properties in quantum systems
\cite{larkin1969quasiclassical, Almheiri:2013hfa, Shenker:2013pqa,
  Shenker:2014cwa, Roberts:2014isa, Maldacena:2015waa, Kitaev:2015aa,
  Maldacena:2016hyu}. The simplest and best studied example is the OTO
four point function, given by the real time thermal correlation
function
\begin{equation}
  \label{eq:oto}
  C(t,x) \equiv \langle W(t,x) V(0) W(t,x) V(0)\rangle_{\beta}\, ,
\end{equation}
with the subscript denoting a thermal expectation value computed at
inverse temperature $\beta = T^{{-}1}$. Here $V$ and $W$ are simple
local operators containing a small number of degrees of freedom,
e.g.~a primary of ${\cal O}(1)$ conformal weight in a conformal field
theory (CFT). This correlation function governs the non-trivial time
dependence of a commutator norm $\langle[W(t,x),V(0)]^2\rangle$ and
can be used to probe a quantum version of the butterfly effect,
demonstrating how a change in initial conditions affects later
measurements. It has been argued that, at least for some classes of
quantum systems, $C(t,x)$ behaves as
\begin{equation}
  C(t,x) \approx c_0 + c_1 \erm^{\lambda_L (t - x/v_B)}\, ,
\end{equation}
where $\lambda_L$ is the quantum Lyaponov exponent and $v_B$ is the
speed characterizing how quickly the Lyaponov growth progagates
through the system, referred to as the butterfly velocity.

Part of the excitement regarding chaos and the Lyaponov exponent
arises from holographic considerations. General properties of thermal
quantum correlation functions lead to an upper bound on $\lambda_L$,
$\lambda_L \leq {2\pi \over \beta}$, with the bound saturated in
holographic systems with black holes present in the bulk
\cite{Maldacena:2015waa}. Since this maximal Lyaponov growth occurs
rather generically in holographic systems in black hole backgrounds,
chaotic properties can be used as a diagnostic of whether a given
conformal field theory admits a simple holographic gravitational
description \cite{Kitaev:2015aa, Perlmutter:2016pkf}, especially in
combination with other known criteria such as a large number of
degrees of freedom and a sparse spectrum of low-lying operators
\cite{Heemskerk:2009pn, ElShowk:2011ag}. It is currently unknown what
the list of necessary and sufficient conditions a CFT must satisfy in
order to have a semiclassical Einstein gravity dual, so any hints
along these lines are welcome.

The downside of the OTOC as a diagnostic of chaos is that it is a
somewhat unfamiliar observable, as traditional field theory
observables tend to be either time-ordered or response functions. In
the context of conformal field theories, the four-point function is
famously the first correlation function whose form is not completely
specified by conformal symmetry, instead requiring a conformal block
decomposition depending explicitly on the OPE coefficients for the
operators under consideration. In addition, the OTOC of interest is a
thermal correlation function, which further exacerbates computational
difficulties (especially in higher dimensions). Controlled
calculations of such thermal conformal block expansions, which must
then be analytically continued to real time, present considerable
technical challenges, even in two dimensions \cite{Roberts:2014ifa,
  Hampapura:2018otw, Liu:2018iki, Chang:2018nzm}.

A very intriguing recent development has been evidence of a relation
between chaotic and hydrodynamic properties, referred to as pole
skipping, in the context of maximally chaotic systems such as
holographic models and SYK models \cite{Blake:2017ris, Blake:2018leo,
  Blake:2019otz}. It has been observed in these and related models
that the thermal energy density two-point functions possess subtle
analyticity properties in the complex frequency and momentum planes
that permit one to extract the Lyaponov exponent and butterfly
velocity. While energy density response functions outside the
hydrodynamic regime are still very complicated observables, the
possible connection between chaos and hydrodynamics is very enticing
and has been further explored in, e.g.~\cite{Grozdanov:2017ajz,
  Grozdanov:2018kkt, Guo:2019csw, Grozdanov:2019uhi, Natsuume:2019xcy,
  Natsuume:2019sfp, Ahn:2019rnq, Natsuume:2019vcv, Wu:2019esr,
  Ceplak:2019ymw, Abbasi:2019rhy, Liu:2020yaf, Ahn:2020bks,
  Abbasi:2020ykq, Jansen:2020hfd, Grozdanov:2020koi}. More precisely,
the pole skipping behavior is revealed by writing the energy density
retarded two-point function as
\begin{equation}
  \label{eq:energy-density}
  G^R_{T^{tt}T^{tt}}(\omega,k) = {N(\omega,k) \over D(\omega,k)}\, ,
\end{equation}
with both $N(\omega,k)$ and $D(\omega,k)$ possessing a line of zeroes
that passes through the point
\begin{align}
  \label{eq:chaos-pt}
  \omega_* ={}& \ii \lambda_L\, , & k_* ={}& \ii k_0 =  \ii{\lambda_L \over v_B}\, .
\end{align}
While the small $\omega$, $k$, behavior of the denominator
$D(\omega,k)$ is determined by hydrodynamic considerations, the
conjecture is that this line of hydrodynamic zeroes analytically
continues into the complex frequency and momentum plane precisely
through the `chaotic' point \eqref{chaos-pt}. This would be pole is
then cancelled by a corresponding zero in the numerator. A simple
consequence of this scenario is that the Green's function directly at
the pole skipping point $(\omega_*, k_*)$ is not uniquely defined and
depends on the direction of approach. This can be seen by expanding
the numerator and denominator near $(\omega_*, k_*)$, yielding
\begin{equation}
  \label{eq:nearby-exp}
  G^R(\omega_*+ \delta \omega, k_* + \delta k) \approx {\partial_\omega N(\omega_*, k_*) (\delta \omega/\delta k) + \partial_k N(\omega_*, k_*) \over \partial_\omega D(\omega_*, k_*) (\delta \omega/\delta k) + \partial_k D(\omega_*, k_*)}\, .
\end{equation}
This ambiguity has been seen to emerge in general holographic settings
from the near-horizon behavior of the relevant wave equations
\cite{Blake:2018leo, Blake:2019otz}, as we briefly review below. In
addition to holographic models, pole skipping has also been observed
in SYK chains and sheds some light on previous observations connecting
chaos and energy diffusion \cite{Gu:2016oyy, Gu:2017ohj, Gu:2017njx}.

In this article, we investigate pole skipping in two dimensional
conformal field theories by considering the behavior of stress tensor
response functions. While traditional hydrodynamics breaks down in two
dimensions, it has been observed in \cite{Haehl:2018izb} that the real
time, thermal stress tensor Green's functions on an infinite spatial
manifold also exhibit pole skipping features analogous to those
present in higher dimensional gravitational systems. However, as noted
in \cite{Haehl:2018izb}, the result is surprising in the sense that
the pole skipping location is universal with any two dimensional CFT
having a `skipped pole' at the position $\omega_* = 2\pi \ii
T$. Identifying this location with the Lyaponov exponent via
$\omega_* = \ii \lambda_L$, one would conclude that every two
dimensional CFT is maximally chaotic, which would indeed be quite
surprising. We revisit this situation and note that by considering a
CFT on a compact spatial manifold, this tension is partially
resolved. The location $\omega_*$ depends on the spectrum of the CFT
through the stress tensor one point function on the torus. Explicitly,
we compute $\langle TT\rangle_{\TT^2}$ on a torus $\TT^2$ with modular
parameter $\tau = {\ii \beta \over 2\pi R}$ (corresponding to inverse
temperature $\beta= T^{{-}1}$ and spatial circle radius $R$) and
analytically continue to find the retarded Green's function. Fourier
transforming, we show that pole skipping in the holomorphic stress
tensor retarded Green's function occurs at
\begin{equation}
  \label{eq:wstar}
  \omega_*^2 = k_*^2 = {24 \over c} \langle T\rangle_{\TT^2}= {-} {12 \over c} {E\over R}\, .
\end{equation}
Here $E$ is the thermodynamic energy, $E={-}\partial_\beta \log Z$, or
alternatively obtained from the local energy density operator via
$\langle T^{tt}\rangle = {E \over 2\pi R}$. Comparing to
\eqref{chaos-pt}, we see that $v_B= \omega_*/k_* = 1$. Upon taking the
high temperature limit, or equivalently the noncompact limit
$R \to \infty$, \eqref{wstar} reproduces the answer obtained in
\cite{Haehl:2018izb} on spatial slice $\RR$, as
${E \over R}\to {c \over 12 } ({2 \pi \over \beta})^2$ by the
universal Cardy asymptotics of unitary two dimensional CFTs.

As a corrollary of our results, we show that, for compact unitary CFTs
with $c>1$, modular invariance implies an upper bound on the energy
density $E/R$ for all temperatures. This immediately gives a bound on
the pole skipping location, which we can write (assuming $\omega_*$ is
purely imaginary, i.e.~$E >0$) as a bound on $\lambda = \Im \omega_*$
\begin{equation}
  \label{eq:ps-bound}
  \lambda \leq 2\pi T\, .
\end{equation}
If we identify the pole skipping location with the Lyaponov exponent
as suggested in \cite{Blake:2017ris}, $\lambda = \lambda_L$, then we
recognize the bound on chaos of \cite{Maldacena:2015waa}. Here we find
the bound as a consequence of modular invariance in compact, unitary
two dimensional CFTs with $c>1$. Furthermore, we note that for any
large $c$ CFT$_2$ with a light sparse spectrum, as quantified by
Hartman, Keller, and Stoica (HKS) in \cite{Hartman:2014oaa}, the
stress tensor one point function is fixed by the free energy of a BTZ
black hole, yielding
${E \over R}= {c \over 12} \left( {2\pi \over \beta}\right)^2$ for all
$\beta < 2\pi R$, and the corresponding pole skipping location
$\omega_* = \ii \lambda$ is precisely the maximal Lyaponov exponent,
$\lambda = 2\pi T$, for all temperatures above the Hawking-Page
transition. We note that there are known large $c$ CFTs with a sparse
light spectrum in this sense that are not expected to have
semiclassical Einstein gravity duals or maximal chaos, such as some
permutation orbifolds \cite{Hartman:2014oaa, Perlmutter:2016pkf,
  Belin:2017jli}, suggesting that the relationship between $\lambda$
and $\lambda_L$ may be more complicated. Nevertheless, we find it
intriguing that, at least in the context of two dimensional CFTs, both
obey the same bound.

The paper is organized as follows. In \secref{diag}, we review some
relevant background, including the gravitational origins of pole
skipping as well as the HKS spectrum constraints required for a
CFT$_2$ to reproduce gravitational thermodynamics in
AdS$_3$. Following this, in \secref{ps-cfts}, we turn to purely CFT
evaluations of the stress tensor response functions, first on the
cylinder, and then moving to the torus. With these results in hand, we
combine the holomorphic and anti-holomorphic results to obtain the
energy density Green's functions and prove the bound
\eqref{ps-bound}. Finally, we conclude and discuss some potential
future directions in \secref{discuss}.

\section{Diagnostics of holographic CFTs}
\label{sec:diag}
In this section, we review some salient features of holographic
conformal field theories that will be relevant for our
analysis. First, we sketch the gravitational arguments for pole
skipping in holographic response functions, where the peculiar
non-analyticities emerge as a consequence of the near-horizon geometry
in black hole backgrounds. Following this, we turn to two-dimensional
conformal field theories and review the constraints imposed on the
spectrum by bulk thermodynamic considerations.

\subsection{Pole skipping}
\label{sec:pole-skip}

Pole skipping is a feature of holographic response functions that
arises due to the nature of the relevant wave equations in the
presence of a black hole horizon. Here we will briefly sketch the
argument for AdS-Schwarzschild black holes in $d \geq 3$
Einstein-Hilbert gravity with negative cosmological constant. Note
however, that due to the lack of propagating gravitational waves in
three dimensions, the analysis is not directly relevant for the BTZ
geometries dual to finite temperature two-dimensional CFTs; in a
sense, the results of section \secref{ps-cfts} will be to demonstrate
that pole skipping does arise in this setting as well.

We consider the real time stress tensor response function
$G_{T^{tt}T^{tt}}^R(\omega, k)$ in an AdS-Schwarzschild black hole in
$d+1 \geq 4$ dimensions. In ingoing Eddington-Finkelstein coordinates,
the metric for such a planar black hole reads
\begin{equation}
  \label{eq:ads-bh}
  \dd s^2 = {L^2 \over r^2} \left[ {-} f(r) \dd v^2 + 2 \dd v \dd r + \dd x_i \dd x^i \right]\, .
\end{equation}
Here the conformal boundary is located at $r=0$, the emblackening
factor is $f(r) = 1 - (r/r_+)^d$, $v = t - r_*$ is the ingoing
Eddington-Finkelstein coordinate (with $r_*$ determined by
$\dd r_* = \dd r/f(r)$), and the boundary spatial coordinate indices
run over $i=1, \dotsc, d-1$. The temperature of the solution is then
$T = {d \over 4\pi r_+}$.

For our purposes, we do not need to obtain the full Green's function
$G^R_{T^{tt}T^{tt}}(\omega,k)$, and it will suffice to illustrate
that, for a particular value of $\omega$ and $k$, the holographic
calculation of such a response function exhibits the novel features
mentioned in the introduction. Recall that to obtain such a response
function, the holographic dictionary instructs us to consider
perturbations of the background \eqref{ads-bh} that solve the
linearized Einstein equations and satisfy ingoing boundary conditions
at the horizon \cite{Son:2002sd}. The response function can then be
read off from this linearized solution by taking the ratio of the
coefficients in an expansion near the conformal boundary ($r=0$). Our
strategy, following very closely the discussion in
\cite{Blake:2018leo}, is to demonstrate how, in constructing a
solution to the linearized Einstein equation via an expansion about
the horizon $r=r_+$, one finds that at the particular point
\begin{align}
  \label{eq:adsbh-wstar}
  \omega_* ={}& 2 \pi \ii T \, , & k_*^2 ={}& {-} {8 \pi^2 T^2 (d-1)\over d}\, ,
\end{align}
an additional ingoing solution emerges and leads to ambiguities in the
response function at this point. More examples of pole skipping in
holographic contexts and their implications can be found in
\cite{Blake:2019otz}.

To see this, we consider the Einstein equation
\begin{equation}
  \label{eq:EE}
  0 = R_{ab} + {d \over L^2} g_{ab}\, ,
\end{equation}
linearized about the solution \eqref{ads-bh}. Decomposing the
perturbations in Fourier modes,
$\delta g_{ab}(v,r,x^i) = \erm^{{-}\ii (\omega v - k x)} \delta
g_{ab}(r)$, where we take the wave vector to point along
$x^1 \equiv x$, we obtain a set of coupled second order differential
equations for the $\delta g_{ab}(r)$. For generic $\omega$ and $k$,
these equations admit two independent solutions, one of which is
ingoing and the other outgoing. We distinguish between the two
solutions by imposing regularity at the horizon. In particular, we can
construct the generic solutions via series expansions about the
horizon, and the ingoing solution will have a regular series expansion
about the singular point $r = r_+$
\begin{equation}
  \label{eq:pertexp}
  \delta g_{ab}(r) = \delta g_{ab}(r_+) + {\cal O}(r - r_+)\, .
\end{equation}
In principle, the wave equation will fix all of the coefficients in
such an expansion in terms of data on the horizon. To see why
\eqref{adsbh-wstar} is a special point, we consider the behavior of
the $vv$ component of \eqref{EE} near the horizon, which reads
\begin{multline}
  \label{eq:vv-nh}
  0 = \left(k^2 - {4 \pi T (d-1) \over d} \ii \omega \right) \delta g_{vv}(r_+) + \left(\omega - 2\pi \ii T \right) \left[2 k \delta g_{vx}(r_+) + \omega \delta g_{x^i x^i}(r_+) \right] \\ + {\cal O}(r-r_+)\, .
\end{multline}
For generic $\omega$ and $k$, this equation imposes a non-trivial
constraint on the initial data $\delta g_{ab}(r_+)$. However,
precisely when $\omega = \omega_* = 2\pi \ii T$, this equation
simplifies dramatically and reduces, for generic $k$, to the
constraint $\delta g_{vv}(r_+) = 0$. If we further tune $k \to k_*$,
then $\delta g_{vv}(r_+)$ is completely unconstrained, and it appears
the generic ingoing solution will have an additional free parameter at
this point. Further analysis of the remaining components of the
linearized equations do not resolve this ambiguity and confirms that
indeed at this point there is an additional ingoing solution with no
outgoing solution.

This additional ingoing solution is the gravitational origin of the
pole skipping phenomena advertised above. While we won't explicitly go
through the details, having this additional solution allows one to
independently tune the coefficients of the near boundary expansion and
hence obtain any value desired for the dual response function. We have
only considered the simplest setting for this behavior, namely pure
Einstein-Hilbert gravity without matter, but the mechanism holds far
more generally \cite{Blake:2018leo}; furthermore, one can find
additional isolated locations in the complex $(\omega,k)$ plane that
admit an extra ingoing solution, both for the stress-tensor
correlators considered here as well as more general scalar and current
response functions \cite{Blake:2019otz}. All of these locations
present non-trivial constraints on the behavior of the dual response
functions, and the crucial ingredient in their emergence is the
presence of a black hole horizon. Therefore, one expects that such
behavior should be considered as a necessary ingredient in a
holographic CFT, and indeed our goal will be to formulate a sharp
constraint in CFT$_2$ for pole skipping precisely at
\eqref{adsbh-wstar} with $d=2$.

\subsection{HKS sparse spectrum}
\label{sec:hks}

We now briefly review the constraints gravitational thermodynamics
imposes on a holographic CFT$_2$, first elucidated in
\cite{Hartman:2014oaa}. The fundamental physical input is the
universal free energy of three-dimensional gravity with a negative
cosmological constant as a function of temperature, with a
Hawking-Page transition from a thermal AdS solution to the BTZ black
hole as the temperature is increased \cite{Hawking:1982dh}. The upshot
of the analysis is a sharp formulation of the notion of a sparse
spectrum of light operators that has long been suspected to be
necessary for a conformal field theory to have a simple holographic
dual \cite{Heemskerk:2009pn, ElShowk:2011ag}. We will see in
\secref{t2} how this universal thermodynamics also gives rise to pole
skipping at a frequency corresponding to a maximal Lyaponov exponent
for high temperatures.

The thermodynamics of three-dimensional gravity in the semiclassical
limit ($c = {3 \ell \over 2 G_N} \to \infty$) can be seen by comparing
the saddle points of the Einstein-Hilbert action in the canonical
ensemble. The saddle points of interest are obtained by modular
transformations of the thermal gas solution in AdS$_3$, which
corresponds to a partition function given solely by the vacuum
character of the dual CFT\footnote{Here we've included the
  contribution of the boundary gravitons generated by the two
  asymptotic Virasoro algebras \cite{Brown:1986nw}. A slogan to
  remember these results is that the gravitational answers are often
  `vacuum dominated', i.e.~given solely by the vacuum block (in an
  appropriate channel) in the dual CFT. For the case at hand, the
  relevant conformal blocks are the Virasoro characters.}:
\begin{equation}
  \label{eq:ZthermAdS}
  Z_{tAdS}(\tau, \bar \tau) = \chi_0(\tau) \bar \chi_0(\bar \tau)\, .
\end{equation}
Here $\chi_0(\tau)$ is the Virasoro vacuum character
\begin{equation}
  \label{eq:chi0}
  \chi_0(\tau) = {(1-q) q^{{-}{c -1 \over 24}} \over \eta(\tau)} = {q^{{-}{c \over 24}} \over \prod_{n=2}^\infty (1-q^n)}\, ,
\end{equation}
where $\eta(\tau)$ is Dedekind $\eta$ function and
$q = \erm^{2\pi \ii \tau}$. For a CFT at inverse temperature
$\beta=T^{{-}1}$ and on a spatial circle of radius $R$, the modular
parameter $\tau$ is given by $\tau = {\ii \beta \over 2\pi R}$, with
$\bar \tau = \tau^*$. This partition function leads to a free energy
of
\begin{equation}
  \label{eq:FtAdS}
  {-} \beta F_{tAdS} = {c \beta\over 12R}  + {\cal O}(1) \, ,
\end{equation}
in the $c \to \infty$ limit. The BTZ free energy is obtained
similarly, by starting with a partition function given by the modular
$S$ transform of the vacuum:
\begin{equation}
  \label{eq:ZBTZ}
  Z_{BTZ}(\tau, \bar \tau) = \chi_0({-} 1/\tau) \chi_0({-} 1/\bar \tau)\, ,
\end{equation}
which yields the free energy
\begin{equation}
  \label{eq:FBTZ}
  {-} \beta F_{BTZ} = {\pi^2 c R\over 3\beta} + {\cal O}(1)\, .
\end{equation}
Comparing these two free energies, we find a phase transition at
$\beta= 2\pi R$, with the black hole solution dominating at high
temperatures. All other modular images turn out to be subleading and
we will not consider them here. 

As demonstrated in \cite{Hartman:2014oaa}, general unitary large $c$
CFT$_2$s need not exhibit this thermodynamic behavior. However, by
studying modular invariance, \cite{Hartman:2014oaa} found that a bound
on the density of light states
\begin{equation}
  \label{eq:HKSbound}
  \rho(h,\bar h) \lesssim \erm^{4\pi \sqrt{h \bar h}}
\end{equation}
is necessary and sufficient condition to reproduce the aforementioned
gravitational thermodynamics. Here $h$ and $\bar h$ are the left and
right conformal weights (eigenvalues of $L_0$ and $\bar L_0$), with a
`light state' being quantified as those satisfying
$h, \bar h < {c \over 24} + \epsilon$ for some small positive
$\epsilon$.

We emphasize that \eqref{HKSbound} is sufficient only for matching the
gravitational thermodynamic behavior, and in particular is not
sufficient to guarantee a simple graviational dual. There are examples
that satisfy \eqref{HKSbound} and yet are not expected to have
Einstein duals, e.g.~permutation orbifolds \cite{Keller:2011xi,
  Hartman:2014oaa, Belin:2014fna, Haehl:2014yla, Belin:2015hwa}.

For our purposes, we will primarily be interested in the high
temperature/BTZ regime $\beta<2\pi R$, where, assuming
\eqref{HKSbound} is satisfied, any putative dual geometry should
contain a black hole in the large $c$ limit. As the presence of a
black hole horizon is crucial both for a non-trivial Lyaponov exponent
\cite{Shenker:2014cwa,Shenker:2013pqa} as well as the pole skipping
discussed in the previous subsection, this is the regime where one
expects to find chaotic behavior in the dual CFT. The goal of the
remainder of this note is to verify this expectation in the energy
density response functions.

\section{Pole skipping in CFT$_2$}
\label{sec:ps-cfts}

We finally turn to the analysis of the stress tensor response
functions in CFT$_2$.\footnote{For the entirety of this paper, we only
  consider conformal field theories with equal holomorphic and
  anti-holomorphic central charges $c=\bar c$.} We first present a
calculation of the relevant Green's function when the spatial manifold
is noncompact, before turning to the more interesting case of a
compact spatial slice. The strategy is to start with the Euclidean two
point function $G^E(\tau_E, x)$, and then analytically continue the
result to obtain the retarded real time two point function $G^R(t,x)$,
via
\begin{equation}
  \label{eq:GR-def}
  G^R(t,x) = {-} \ii \theta(t) \left[G^E(\epsilon + \ii t, x) - G^E({-}\epsilon + \ii t, x) \right]\, ,
\end{equation}
where $\epsilon>0$ is an infinitesimal positive real number. It is
then a matter of Fourier transforming to determine the location of
pole skipping. 

\subsection{$\langle TT\rangle$ on $\RR \times \mathrm{S}^1$}
\label{sec:rs1}

We start with a noncompact spatial slice, so the Euclidean geometry is
$\RR \times \mathrm{S}^1_\beta$, where the radius of the Euclidean
time circle is $\beta$. This result has been previously obtained and
discussed at length in \cite{Haehl:2018izb}, but we include a
presentation here that easily generalizes to the case of a compact
spatial manifold. As is well known, correlation functions on the
cylinder are determined by the corresponding correlation functions on
the complex plane $\CC$, as they are related by a conformal
transformation. Therefore, we first start on the plane, with standard
metric $\dd s^2 = \dd z \dd \bar z$, where the holomorphic stress
tensor has the two point function
\begin{equation}
  \label{eq:TT-plane}
  \left\langle T(z) T(0) \right\rangle_{\CC} = {c \over 2 z^4}\, .
\end{equation}
To obtain the correlation function on $\RR \times \mathrm{S}^1_\beta$,
we use the coordinate transformation $z = \erm^{{2\pi \over \beta} w}$,
where $w=x + \ii \tau$ is a complex coordinate on
$\RR \times \mathrm{S}^1_\beta$. Using the transformation law for the
stress tensor, we have
\begin{equation}
  \label{eq:T-trans}
  T_{\RR\times \SSS^1}(w) = \left( {2\pi \over \beta} \right)^2 \left[z^2 T_{\CC}(z) - {c \over 24} \right]\, ,
\end{equation}
and so the connected correlation function on the cylinder reads
\begin{align}
  \label{eq:TT-cyl}
  G^E_{\RR \times \SSS^1}(\tau, x) \equiv{}& \left\langle T(w) T(0) \right\rangle_{\RR \times \SSS^1} - \langle T\rangle_{\RR \times \SSS^1}^2 = \left( {2\pi\over \beta} \right)^4 \erm^{{4\pi \over \beta} w} \left\langle T_{\CC}\left(\erm^{{2\pi \over \beta} w}\right) T_{\CC}(1)\right\rangle \nonumber \\
  ={}& {c \over 32} \left({2\pi \over \beta} \right)^4 \sinh^{{-}4}(\pi w/\beta)\, .
\end{align}

We now analytically continue to find the real time Green's
function. This amounts to setting $\tau = \epsilon+ \ii t$, where the
sign of $\epsilon$ determines the operator ordering\footnote{For a
  discussion of the appropriate $\epsilon$ prescriptions for
  Lorentzian correlators, see e.g.~the recent works
  \cite{Roberts:2014ifa, Hartman:2015lfa} or the textbook accounts in
  \cite{streater2000pct, haag2012local}.}. The retarded Green's
function is defined as the commutator
\begin{equation}
  \label{eq:retdef}
  G^R(t,x) = {-} \ii \theta(t) \left\langle \left[{\cal O}(t,x), {\cal O}(0,0) \right]\right\rangle\, ,
\end{equation}
where $\theta(t)$ is the Heaviside theta function. For the case at
hand we find that the stress tensor response function is
\begin{align}
  \label{eq:TT-ret}
  G^R_{\RR \times \SSS^1}(t, x) ={}& {-} \ii\theta(t) \left[G^E_{\RR \times \SSS^1}(\epsilon + \ii t, x) - G^E_{\RR \times \SSS^1}({-} \epsilon + \ii t, x) \right] \\
  ={}& {-} {\ii c \over 32} \left({2 \pi \over \beta} \right)^4 \theta(t) \left[ \sinh^{{-}4}\left({\pi \over \beta} (x - t+ \ii \epsilon) \right) - \sinh^{{-}4}\left({\pi \over \beta} (x - t - \ii \epsilon) \right)\right]\, . \nonumber
\end{align}

\begin{figure}[t]
  \centering
  \includegraphics{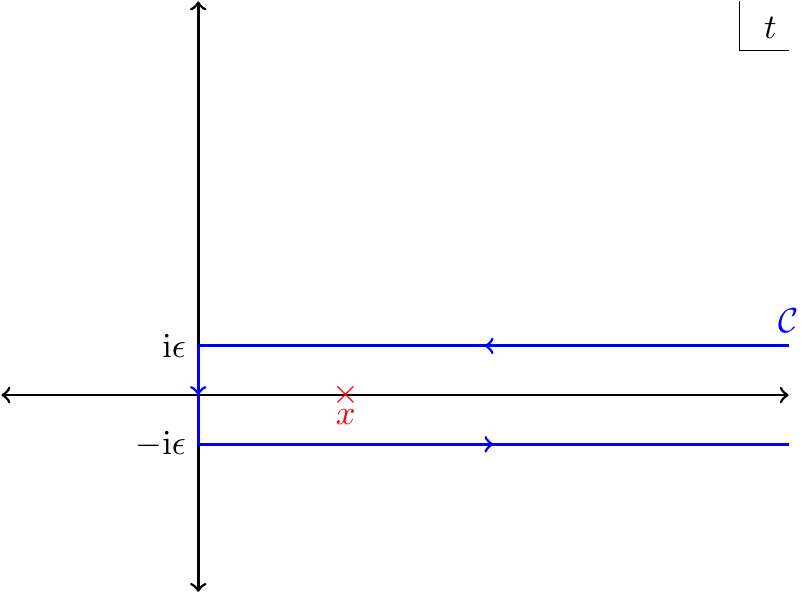}
  \caption{Contour ${\cal C}$ in the complex (Lorentzian) time plane
    used in evaluating the Fourier transform of
    $G^R_{\RR\times \SSS^1}(t,x)$, shown with $x>0$. The two branches,
    infinitessimally close to the real line, correspond to the two
    terms in the commutator defining $G^R$. The Heaviside function in
    the definition of $G^R$ restricts the contour to positive real
    times, and hence the singularity at $t=x$ is not enclosed by the
    contour of integration for $x<0$.}
  \label{fig:RS1fig}
\end{figure}

All that remains is to Fourier transform this result. In fact, the
detailed form the retarded two point function is not terribly
important; we can rewrite the Fourier transform of $G^R$ as a contour
integral of the Euclidean Green's function, requiring only knowledge
of its poles and their residues. Performing the integral over time
first, we see that
\begin{align}
  \label{eq:GRFT}
  G^R_{\RR \times \SSS^1}(\omega, x) ={}& \int_{{-} \infty}^\infty \dd t\, \erm^{\ii \omega t} G^R_{\RR \times \SSS^1}(t, x) = {-} \ii \int_0^\infty \dd t\, \erm^{\ii \omega t}  \left[G^E_{\RR \times \SSS^1}(\epsilon + \ii t, x) - G^E_{\RR \times \SSS^1}({-} \epsilon + \ii t, x) \right] \nonumber \\
  ={}& {-} \ii \int_{{\cal C}} \dd t \, \erm^{\ii \omega t} G^E_{\RR \times \SSS^1}(\ii t, x)\, ,
\end{align}
where ${\cal C}$ is a contour that wraps the positive real $t$-axis
counterclockwise. To see this, note that we can shift the integration
variable for the first term in the square brackets to
$t \to t - \ii \epsilon$, and similarly shift the second term
$t \to t + \ii \epsilon$ to obtain the integral
\begin{align}
  \left\{\erm^{{-}\omega \epsilon} \int_{{-}\ii \epsilon}^{\infty - \ii \epsilon}\dd t - \erm^{\omega \epsilon} \int_{\ii \epsilon}^{\infty + \ii \epsilon} \dd t\right\} \erm^{\ii \omega t} G^E_{\RR \times \SSS^1}(\ii t, x)\, .
\end{align}
As we send $\epsilon \to 0$, we land on the contour ${\cal C}$ shown
in \figref{RS1fig}.

From this contour manipulation, we have reduced the calculation to the
evaluation of a residue. From \eqref{TT-cyl}, we see that
$G^E_{\RR \times \SSS^1}(\ii t, x)$ has a pole when $x = t$, which is
enclosed by ${\cal C}$ when $x>0$, and we find a residue of
\begin{equation}
  2\pi \ii \Res_{t=x} \left[\erm^{\ii \omega t} G^E_{\RR \times \SSS^1}(\ii t, x) \right] = {\pi^2 c \over 6} \theta(x) \erm^{\ii \omega x} \omega \left[\omega^2 + \left({2\pi \over \beta} \right)^2 \right]\, .
\end{equation}
Finally we can Fourier transform in $x$ (including an infinitesimal
positive imaginary component to $\omega$ for convergence purposes) to
find \cite{Haehl:2018izb}
\begin{equation}
  \label{eq:GRwk-cyl}
  G^R_{\RR \times \SSS^1}(\omega,k) = {\pi c \over 6} { \omega \left[\omega^2 + \left({2\pi \over \beta} \right)^2 \right] \over \omega - k}\, .
\end{equation}

With this result, we see that any conformal field theory on the
cylinder in two dimensions exhibits pole skipping at
$\omega_* = k_* = 2\pi \ii T$, in agreement with the gravitational
result \eqref{adsbh-wstar} in higher dimensions continued to
$d=2$. Identifying the pole skipping location with Lyaponov exponent,
one arrives at the conclusion that all two dimensional CFTs are
maximally chaotic, which seems counterintuitive. We next turn to the
analogous calculation on a torus, with a compact spatial manifold,
where we will see that this situation is at least partially remedied.

\subsection{$\langle TT \rangle$ on $\TT^2$}
\label{sec:t2}

To perform the same calculation on the torus, we use the fact that the
stress tensor two-point function on the torus is also completely fixed
by Virasoro Ward identities, up to the value of the one point function
$\langle T\rangle_{\TT^2}$. More explicitly, the two-point function of
interest is given in terms of the Weierstrass elliptic function
$\wp(z)$ by \cite{Eguchi:1986sb}
\begin{align}
  \label{eq:TT-T2}
  G^E_{\TT^2}(z-w) \equiv{}& \left\langle T(z) T(w)\right\rangle_{\TT^2} - \langle T\rangle^2_{\TT^2} \nonumber \\
  ={}& {c \over 12} \wp''(z-w) + 2 \left[ \wp(z-w) + 2\eta_1 \right] \langle T \rangle_{\TT^2} + 2 \pi \ii \partial_\tau \langle T \rangle_{\TT^2}\, .
\end{align}
The Weierstrass elliptic function is defined as
\begin{equation}
  \label{eq:weierstrass}
  \wp(z) = {1 \over z^2} + \sum_{(m,n) \neq (0,0)} \left[{1 \over (z + 2\pi R m + \ii n \beta)^2} - {1 \over (2\pi R m + \ii n \beta)^2}\right]\, ,
\end{equation}
and $\eta_1 = \zeta(1/2)$ is a constant, where
$\zeta'(z) = {-} \wp(z)$. The modular parameter of the torus $\tau$ is
given by $\tau = {\ii \beta \over 2\pi R}$, with $R$ the radius of our
spatial circle. A simple consistency check of \eqref{TT-T2} is to note
that it reproduces the correct behavior in the OPE limit $z \to w$,
i.e.~${c \over 2(z -w)^4} + {2 \langle T \rangle \over (z-w)^2}$. The
stress tensor one point function is determined by the partition
function using\footnote{Note that the factor of $R^2$ and a sign in
  this expression result from a different map from the plane to the
  cylinder compared to the previous subsection. Here we take the plane
  coordinate to be given by $z=\erm^{{-}\ii (x+\ii \tau_E)/R}$ with
  $x + \ii \tau_E$ the complex coordinate on this more standard radial
  quantization cylinder. This choice also leads to a relative sign
  between the energy density $T_{\tau_E\tau_E}$ and $T+\bar T$.}
\begin{equation}
  \label{eq:Tvev}
  \langle T\rangle_{\TT^2} = {\ii \over 2\pi R^2} \partial_\tau \log Z\, .
\end{equation}

We can now follow the same strategy as the previous subsection,
writing $z = x + \ii \tau_E$, using a subscript $E$ to distinguish
Euclidean time from the torus modular parameter, and evaluating the
real time retarded Green function via
\begin{equation}
  \label{eq:T2-GR}
  G^R_{\TT^2}(t,x) = {-} \ii\theta(t) \left[G^E(x, \epsilon + \ii t) - G^E(x, {-} \epsilon + \ii t) \right]\, ,
\end{equation} 
with $G^E(x,\tau_E) = G^E(x + \ii \tau_E)$. Again, the two terms give
the two terms in a commutator $[T(x,t), T(0)]$ thanks to the
$\epsilon$ prescription. Conveniently, many of the constant terms in
the Euclidean Green's function drop out in the difference, and so we
are left with
\begin{equation}
  \label{eq:T2-GR2}
  G^R_{\TT^2}(t,x) = {-} \ii \theta(t) \left[ {c \over 12} \wp''(x - t + \ii \epsilon) + 2 \langle T\rangle_{\TT^2} \wp(x - t + \ii \epsilon) - (\epsilon \to {-} \epsilon) \right]\, .
\end{equation}

\begin{figure}[t]
  \centering
  \includegraphics{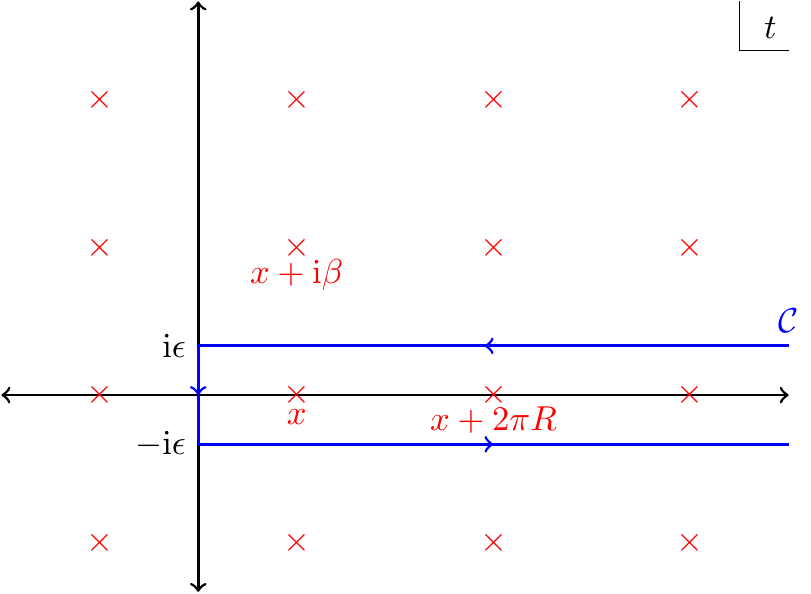}
  \caption{Contour used in evaluating the Fourier transform of
    $G^R_{\TT^2}$. Here the singularity at $t=x$ has doubly periodic
    images, separated by $\ii \beta$ and $2\pi R$, as required by the
    torus geometry.}
    \label{fig:T2fig}
\end{figure}

To perform the Fourier transform, we perform the same manipulations we
did on the cylinder, ending up with a contour integral of the form
\begin{equation}
  \label{eq:GRT2FT}
  G^R_{\TT^2}(\omega,x) = {-} \ii \int_{{\cal C}} \dd t \, \erm^{\ii \omega t} \left[  {c \over 12} \wp''(x - t) + 2 \langle T\rangle_{\TT^2} \wp(x - t) \right]\, .
\end{equation}
We see that the only difference between the torus and the cylinder is
that now an infinite series of poles contribute, thanks to the images
of the OPE singularities on the compact spatial slice. See
\figref{T2fig} for an illustration. The remaining integrals are
straightforward to evaluate using \eqref{weierstrass}, and one finds
\begin{align}
  \int_{{\cal C}} \dd t\, \erm^{\ii \omega t} \wp(x-t) ={}& {-} 2 \pi \omega \sum_{x + 2\pi m R>0} \erm^{\ii \omega(x+2\pi R m)}\, , \\
  \int_{{\cal C}} \dd t\, \erm^{\ii \omega t} \wp''(x-t) ={}& 2 \pi \omega^3 \sum_{x + 2\pi m R>0} \erm^{\ii \omega(x+2\pi R m)}\, .
\end{align}
These are geometric series and, provided we add an infinitesimal
positive imaginary part to $\omega$, we can take $0 < x < 2\pi R$ and
evaluate the sums over $m \geq 0$ to find
\begin{equation}
  \label{eq:GRT2FT2}
  G^R_{\TT^2}(\omega,x) = {-} {\ii c \over 12} {2\pi \omega \erm^{\ii \omega x} \over 1 - \erm^{2\pi \ii R \omega}} \left[ \omega^2 - {\langle T\rangle_{\TT^2} \over c/24} \right]\, .
\end{equation}
Performing the last Fourier transform over $x$, we obtain
\begin{equation}
  \label{eq:GRT2final}
  G^R_{\TT^2}(\omega,k_n) = {\pi c \over 6} {\omega \over \omega - k_n} \left[ \omega^2 - {\langle T\rangle_{\TT^2} \over c/24} \right]\, .
\end{equation}
Here $k_n = n/R$ is the discrete Fourier momentum on a compact spatial
circle.

With this result, we see that the location of the zero in the
numerator of the response function is determined by
$\langle T\rangle$, and so the pole skipping location moves to
$\omega_*^2 = k_*^2 = {\langle T\rangle \over (c/24)}$. Note that this
result reduces to that obtained in the previous subsection as we send
$R\to \infty$. This is because in the non-compact limit, which is
equivalent to the high-temperature limit $\beta \to 0$ by modular
invariance, the partition function, and hence $\langle T\rangle$, for
every unitary CFT$_2$ will be dominated by the modular $S$ transform
of the Virasoro vacuum character, leading to a value of
$\langle T\rangle = {-} c/24 (2\pi/\beta)^2$.

If we now consider CFTs that satisfy the HKS spectrum condition
\eqref{HKSbound}, then the universal high temperature free energy
leads to
\begin{equation}
  \label{eq:THKS}
  \langle T\rangle_{HKS} = {-} {c \over 24} \left( {2\pi \over \beta} \right)^2\,,
\end{equation}
for all $\beta < 2\pi R$, so that the pole skipping occurs at
$\omega_* = k_* = 2\pi \ii T$ for all sufficiently high
temperatures. Thus we conclude that imposing the HKS condition is
sufficient to guarantee that pole skipping occurs at the `maximal
chaos' location for all temperatures above the Hawking-Page transition
in the gravitational dual, as expected by the presence of a black
hole.

\subsection{Energy density pole skipping and a bound on $\omega_*$}
\label{sec:bound}

Thus far we have focused exclusively on the behavior of the
holomorphic stress tensor response functions. For the honest energy
density response functions, we use
${-}2\pi T_{\tau\tau} = T + \bar T$, and so the connected Euclidean
Green's function schematically reads
\begin{equation}
  \label{eq:Ttt}
  \langle T_{\tau \tau}T_{\tau\tau}\rangle^{c} = \langle TT\rangle^c + \langle \bar T \bar T\rangle^c\, .
\end{equation}
The antiholomorphic contribution can be derived by exactly the same
procedure as $\langle TT\rangle$ above, and results in the simple
substitution $k \to {-} k$. The end result after combining these two
pieces is
\begin{equation}
  \label{eq:Tttfinal}
  G^R_{T^{tt}T^{tt}}(\omega,k) = {c \over 12\pi} {\omega^2 \over \omega^2 - k^2} \left[\omega^2 + {2\pi\langle T^{tt}\rangle \over c/12} \right] = {c \over 12\pi} {\omega^2 \over \omega^2 - k^2} \left[\omega^2 + {12 \over c} {E \over R} \right]\, .
\end{equation}
Since we have $\bar \tau = \tau^*$ and no angular potential,
$\bar \tau = {-} \tau$, we have dropped a term proportional to
$\langle T - \bar T\rangle$, as
$\langle T\rangle = \langle \bar T\rangle$.  We have also used the
fact that $T^{tt}$ is the local energy density, and so
$2\pi \langle T^{tt}\rangle = {E \over R}$, where
$E = {-} \partial_\beta \log Z$ is the thermodynamic energy. Thus we
see that the energy density also exhibits pole skipping with the
location set by $\langle T^{tt}\rangle$,
$\omega_*^2 = {-} {12 \over c} {E \over R}$. Evaluating for a CFT with
a sparse light spectrum in the high temperature/BTZ regime, where
${E \over R} = {\pi^2 c\over 3\beta^2}$, again yields
$\omega_*^2 = {-} (2\pi/\beta)^2.$

As a corollary of this result, we note that the pole skipping
location, assuming $\omega_*$ is purely imaginary, monotonically moves
upward in the upper half plane as the temperature is increased, since
$\partial_T E \geq 0$. This is simply a consequence of thermodynamics,
as $\partial_T E$ is the heat capacity of the system and must be
positive for stability of the ensemble.

In fact, we can learn more using modular invariance. We claim that the
energy density is bounded for all temperatures,
\begin{equation}
  \label{eq:Ebound}
  {E \over R} \leq {\pi^2 c \over 3 \beta^2}\, .
\end{equation}
in any unitary two dimensional CFT with a discrete spectrum and
$c>1$. This directly leads to a bound on $\lambda = \Im \omega_*$
\begin{equation}
  \label{eq:lambdabound}
  \lambda \leq 2 \pi T\, .
\end{equation}
We note that while Cardy asymptotics imply that
${E \over R} \to {\pi^2 c \over 3 \beta^2}$ as $\beta \to 0$, it is
not immediately obvious to us that such a bound must hold for all
$T$. Fortunately, the bound is a simple consequence of modular
invariance, at least for $c>1$. To see this, we write the partition
function in terms of the Virasoro characters
\begin{align}
  \label{eq:characters}
  \chi_0(\tau) ={}& {(1-q)q^{{-}{c-1 \over 24}} \over \eta(\tau)} = {q^{{-}{c \over 24}} \over \prod_{n=2}(1-q^n)}\, , & \chi_h(\tau) ={}& {q^{h-{c-1\over 24}} \over \eta(\tau)} = {q^{h-{c \over 24}} \over \prod_{n=1}(1-q^n)} \, .
\end{align}
Here it is important we are working at $c>1$ in order to ensure the
absence of null states. Evaluating the partition function at
$\tau = {-} \bar \tau = {\ii \beta \over 2 \pi R}$ yields
\begin{equation}
  \label{eq:partition}
  Z(\beta/R) = {\erm^{{\beta c\over 12 R} } \over \prod_{n=2}^\infty (1-\erm^{{-}n \beta \over R})^2} \left[1 + \sum_p {n_p \erm^{{-}\beta \Delta_p \over R} \over \left(1- \erm^{{-}\beta \over R} \right)^2 } \right]\, ,
\end{equation}
where the sum runs over all Virasoro primary operators in the CFT and
$n_p$ is the degeneracy of primaries with dimension
$\Delta_p = h_p + \bar h_p$. Modular invariance tells us that
\begin{equation}
  \label{eq:modinv}
  Z(\beta/R) = Z(4\pi^2 R/\beta)\, .
\end{equation}
Computing the energy $E = {-} \partial_\beta \log Z$ after using this
modular transformation, one finds
\begin{equation}
  \label{eq:CFTenergy}
  {\beta^2 \over 4\pi^2 R} E = {c \over 12} - { \sum_p n_p \erm^{{-}4\pi^2 R \Delta_p \over \beta} \left(\Delta_p + \erm^{{-}2\pi^2 R \over \beta} \csch {2\pi^2 R \over \beta}\right) \over \left(1- \erm^{{-} 4\pi^2 R \over \beta} \right)^2 + \sum_q n_q \erm^{{-} 4\pi^2 R \Delta_q \over \beta}} - \sum_{n=2}^\infty { 2n \over \erm^{4\pi^2 n R \over \beta} - 1}\, .
\end{equation}
While this expression is a bit complicated, all we need is the fact
that the last two terms on the right hand side are always negative,
since the degeneracies $n_p$ are positive integers and the conformal
weights are always positive in a unitary theory, $\Delta_p >
0$. Rearranging, we have
${c \over 12} - {\beta^2\over 4\pi^2} {E \over R} \geq 0$, or
\begin{equation}
  \label{eq:ERbound}
  {12 \over c} {E \over R} \leq \left({2\pi \over \beta} \right)^2\, .
\end{equation}

Combining this with our pole skipping result above, we immediately
find a bound on pole skipping for any compact, unitary CFT$_2$ with
$c>1$: writing $\omega_* = \ii \lambda$, assuming $E>0$, we have
\begin{equation}
  \lambda \leq 2 \pi T\, .
\end{equation}
Identifying $\lambda$ with the Lyaponov exponent $\lambda_L$, we see
that this is none other than the chaos bound of
\cite{Maldacena:2015waa}. As mentioned at the end of the previous
subsection, a sparse light spectrum fixes the value of
$\langle T\rangle$, and hence the energy density, which leads to
\begin{equation}
  \label{eq:EHKS}
  E_{HKS} = 2\pi R \langle T^{tt} \rangle_{HKS} = {-} 2 R \langle T \rangle_{HKS} = {c R \over 12} \left({2\pi \over \beta} \right)^2\, .
\end{equation}
Thus the bound \eqref{ERbound} is saturated and we have maximal pole
skipping for all temperatures $T$ above the self dual point
$2\pi R T = 1$.

For simplicity, here we have only considered the bound \eqref{ERbound}
in theories with $c>1$, to avoid complications due to null states. We
have numerically checked that the bound also holds for the Ising
minimal model, but we leave a systematic check of all Virasoro minimal
models to future work. In addition, we have only considered a system
with no angular potential, i.e.~a purely imaginary modular parameter
$\Re \tau=0$. It is straightforward to determine the pole skipping
location when $\Re \tau \neq 0$, which introduces a dependence on
$\langle T - \bar T\rangle \sim P/R$, where $P$ is the expectation
value of the total momentum on the circle. It seems possible that a
more sophisticated analysis of modular invariance in this setup could
lead to a more general bound on $\lambda$, such as that of
\cite{Halder:2019ric}, but we leave this for future study.\footnote{We
  thank Mukund Rangamani for comments on this point.}

The fact that all CFTs exhibit maximal pole skipping on the cylinder
is a consequence of the fact that we are considering a CFT on an
infinite spatial manifold, which is equivalent by modular invariance
to infinite temperature. Hence the maximal pole skipping on the
cylinder is a consequence of the famous Cardy asymptotics governing
every unitary CFT$_2$. In this section, we've seen that on finite
spatial slices pole skipping is sensitive to the spectrum of the
theory, and that the pole skipping location is bounded by the
temperature in the same manner as the Lyaponov exponent. The
gravitational result, namely that this upper bound is then saturated
for all sufficiently high temperatures on finite spatial slices,
further requires a sparse light spectrum at large central charge, in
the manner of HKS.

\section{Discussion}
\label{sec:discuss}

In this note, we've revisited pole skipping in CFT$_2$. By considering
the behavior of a generic CFT on the torus, we have shown that the
pole skipping location depends on the spectrum of the CFT through the
value of the stress tensor one point function $\langle T\rangle$. This
indicates that at least on a compact spatial manifold the pole
skipping behavior is not universal as suggested by considerations on a
spatial line $\RR$. A corrollary of this result, following from a
bound on the energy density due to modular invariance, indicates that
the pole skipping location for a generic compact unitary CFT$_2$ with
$c>1$ is bounded, $\lambda \leq 2\pi T$, where
$\omega_* = \ii \lambda$. Identifying this location with the Lyaponov
exponent, $\lambda = \lambda_L$, this bound reproduces the MSS chaos
bound of \cite{Maldacena:2015waa}. Furthermore the spectrum
constraints necessary to match the CFT thermodynamics at large central
charge with that of a putative gravitational dual suffice to saturate
this bound, fixing the pole skipping location to that expected from
gravitational calculations in a black hole background, corresponding
to a maximal Lyaponov exponent.

In this work, we have only focussed on the simplest example of pole
skipping, and the work is dramatically simplified by the fact that the
stress tensor correlators are essentially completely fixed by Virasoro
symmetry. However, pole skipping is believed to be a very general
feature of holographic response functions, and furthermore any given
response function can have many different pole skipping locations,
typically lying in the lower half of the complex frequency plane
\cite{Blake:2019otz}. It is likely that requiring similar features in
more non-trivial Green's functions will lead to further constraints on
any dual CFTs. It is well understood how some features of AdS black
holes, for instance quasinormal modes in BTZ \cite{Birmingham:2001pj},
match those obtained directly via CFT calculations on a Euclidean
cylinder; however as we've seen in this note, such matching becomes
more involved once moving beyond thermal response functions on an
infinite spatial manifold. For instance, scalar two point functions on
the torus require conformal block decompositions, and so matching to
gravitational answers, which are much less sensitive to compact
vs.~non-compact spatial manifolds, will likely require constraints on
OPE coefficients, perhaps along the lines of those imposed by higher
genus modular bootstrap considerations \cite{Cho:2017fzo}. It would be
very interesting to see if progress can be made along these
directions.

Furthermore, pole skipping has been conjectured to play an important
role in effective field theories for quantum chaos
\cite{Blake:2017ris, Haehl:2018izb}. For instance,
\cite{Haehl:2018izb} has used the stress tensor cylinder two-point
function in CFT$_2$ to construct an effective action for chaos (with
some aspects generalized to higher dimensions in
\cite{Haehl:2019eae}), where the pole skipping in the upper half plane
produces exponentially growing terms in the propagators of soft modes
in the effective field theory. It is conjectured that the exchange of
these soft modes are responsible for the behavior of OTOCs at large
central charge, with the maximal pole skipping location giving rise to
a maximal Lyaponov exponent in the OTOCs. As we have seen that the
location of pole skipping is in fact sensitive to the spectrum of the
theory, this opens up a window to possibly explore such effective
field theories in a scenario with non-maximal chaos. In addition,
having an independent expectation for the location of pole skipping
suggests that more tests of the relation between pole skipping and
Lyaponov growth in OTOCs could be possible. For instance, perhaps
progress can be made in computing OTOCs in sufficiently simple CFTs
and comparing Lyaponov growth with the corresponding stress tensor
expectation values.

\acknowledgments

The author is grateful to Chi-Ming Chang, Felix Haehl, and Mukund
Rangamani for insightful discussions, and to Mukund Rangamani and Sean
Hartnoll for comments on the draft.


\bibliography{pole-skipping} 

\providecommand{\href}[2]{#2}\begingroup\raggedright\begin{thebibliography}{10}

\bibitem{Hartman:2014oaa}
T.~Hartman, C.A.~Keller and B.~Stoica, \emph{{Universal Spectrum of 2d
  Conformal Field Theory in the Large c Limit}},
  \href{https://doi.org/10.1007/JHEP09(2014)118}{\emph{JHEP} {\bfseries 09}
  (2014) 118} [\href{https://arxiv.org/abs/1405.5137}{{\ttfamily 1405.5137}}].

\bibitem{larkin1969quasiclassical}
A.~Larkin and Y.N.~Ovchinnikov, \emph{Quasiclassical method in the theory of
  superconductivity}, {\emph{Sov Phys JETP} {\bfseries 28} (1969) 1200}.

\bibitem{Almheiri:2013hfa}
A.~Almheiri, D.~Marolf, J.~Polchinski, D.~Stanford and J.~Sully, \emph{{An
  Apologia for Firewalls}},
  \href{https://doi.org/10.1007/JHEP09(2013)018}{\emph{JHEP} {\bfseries 09}
  (2013) 018} [\href{https://arxiv.org/abs/1304.6483}{{\ttfamily 1304.6483}}].

\bibitem{Shenker:2013pqa}
S.H.~Shenker and D.~Stanford, \emph{{Black holes and the butterfly effect}},
  \href{https://doi.org/10.1007/JHEP03(2014)067}{\emph{JHEP} {\bfseries 03}
  (2014) 067} [\href{https://arxiv.org/abs/1306.0622}{{\ttfamily 1306.0622}}].

\bibitem{Shenker:2014cwa}
S.H.~Shenker and D.~Stanford, \emph{{Stringy effects in scrambling}},
  \href{https://doi.org/10.1007/JHEP05(2015)132}{\emph{JHEP} {\bfseries 05}
  (2015) 132} [\href{https://arxiv.org/abs/1412.6087}{{\ttfamily 1412.6087}}].

\bibitem{Roberts:2014isa}
D.A.~Roberts, D.~Stanford and L.~Susskind, \emph{{Localized shocks}},
  \href{https://doi.org/10.1007/JHEP03(2015)051}{\emph{JHEP} {\bfseries 03}
  (2015) 051} [\href{https://arxiv.org/abs/1409.8180}{{\ttfamily 1409.8180}}].

\bibitem{Maldacena:2015waa}
J.~Maldacena, S.H.~Shenker and D.~Stanford, \emph{{A bound on chaos}},
  \href{https://doi.org/10.1007/JHEP08(2016)106}{\emph{JHEP} {\bfseries 08}
  (2016) 106} [\href{https://arxiv.org/abs/1503.01409}{{\ttfamily
  1503.01409}}].

\bibitem{Kitaev:2015aa}
A.~Kitaev, \emph{A simple model of quantum holography.}, {\emph{Talks at KITP,
  April 7, and May 27} (2015) }.

\bibitem{Maldacena:2016hyu}
J.~Maldacena and D.~Stanford, \emph{{Remarks on the Sachdev-Ye-Kitaev model}},
  \href{https://doi.org/10.1103/PhysRevD.94.106002}{\emph{Phys. Rev. D}
  {\bfseries 94} (2016) 106002}
  [\href{https://arxiv.org/abs/1604.07818}{{\ttfamily 1604.07818}}].

\bibitem{Perlmutter:2016pkf}
E.~Perlmutter, \emph{{Bounding the Space of Holographic CFTs with Chaos}},
  \href{https://doi.org/10.1007/JHEP10(2016)069}{\emph{JHEP} {\bfseries 10}
  (2016) 069} [\href{https://arxiv.org/abs/1602.08272}{{\ttfamily
  1602.08272}}].

\bibitem{Heemskerk:2009pn}
I.~Heemskerk, J.~Penedones, J.~Polchinski and J.~Sully, \emph{{Holography from
  Conformal Field Theory}},
  \href{https://doi.org/10.1088/1126-6708/2009/10/079}{\emph{JHEP} {\bfseries
  10} (2009) 079} [\href{https://arxiv.org/abs/0907.0151}{{\ttfamily
  0907.0151}}].

\bibitem{ElShowk:2011ag}
S.~El-Showk and K.~Papadodimas, \emph{{Emergent Spacetime and Holographic
  CFTs}}, \href{https://doi.org/10.1007/JHEP10(2012)106}{\emph{JHEP} {\bfseries
  10} (2012) 106} [\href{https://arxiv.org/abs/1101.4163}{{\ttfamily
  1101.4163}}].

\bibitem{Roberts:2014ifa}
D.A.~Roberts and D.~Stanford, \emph{{Two-dimensional conformal field theory and
  the butterfly effect}},
  \href{https://doi.org/10.1103/PhysRevLett.115.131603}{\emph{Phys. Rev. Lett.}
  {\bfseries 115} (2015) 131603}
  [\href{https://arxiv.org/abs/1412.5123}{{\ttfamily 1412.5123}}].

\bibitem{Hampapura:2018otw}
H.R.~Hampapura, A.~Rolph and B.~Stoica, \emph{{Scrambling in Two-Dimensional
  Conformal Field Theories with Light and Smeared Operators}},
  \href{https://doi.org/10.1103/PhysRevD.99.106010}{\emph{Phys. Rev. D}
  {\bfseries 99} (2019) 106010}
  [\href{https://arxiv.org/abs/1809.09651}{{\ttfamily 1809.09651}}].

\bibitem{Liu:2018iki}
C.~Liu and D.A.~Lowe, \emph{{Notes on Scrambling in Conformal Field Theory}},
  \href{https://doi.org/10.1103/PhysRevD.98.126013}{\emph{Phys. Rev. D}
  {\bfseries 98} (2018) 126013}
  [\href{https://arxiv.org/abs/1808.09886}{{\ttfamily 1808.09886}}].

\bibitem{Chang:2018nzm}
C.-M.~Chang, D.M.~Ramirez and M.~Rangamani, \emph{{Spinning constraints on
  chaotic large $c$ CFTs}},
  \href{https://doi.org/10.1007/JHEP03(2019)068}{\emph{JHEP} {\bfseries 03}
  (2019) 068} [\href{https://arxiv.org/abs/1812.05585}{{\ttfamily
  1812.05585}}].

\bibitem{Blake:2017ris}
M.~Blake, H.~Lee and H.~Liu, \emph{{A quantum hydrodynamical description for
  scrambling and many-body chaos}},
  \href{https://doi.org/10.1007/JHEP10(2018)127}{\emph{JHEP} {\bfseries 10}
  (2018) 127} [\href{https://arxiv.org/abs/1801.00010}{{\ttfamily
  1801.00010}}].

\bibitem{Blake:2018leo}
M.~Blake, R.A.~Davison, S.~Grozdanov and H.~Liu, \emph{{Many-body chaos and
  energy dynamics in holography}},
  \href{https://doi.org/10.1007/JHEP10(2018)035}{\emph{JHEP} {\bfseries 10}
  (2018) 035} [\href{https://arxiv.org/abs/1809.01169}{{\ttfamily
  1809.01169}}].

\bibitem{Blake:2019otz}
M.~Blake, R.A.~Davison and D.~Vegh, \emph{{Horizon constraints on holographic
  Green's functions}},
  \href{https://doi.org/10.1007/JHEP01(2020)077}{\emph{JHEP} {\bfseries 01}
  (2020) 077} [\href{https://arxiv.org/abs/1904.12883}{{\ttfamily
  1904.12883}}].

\bibitem{Grozdanov:2017ajz}
S.~Grozdanov, K.~Schalm and V.~Scopelliti, \emph{{Black hole scrambling from
  hydrodynamics}},
  \href{https://doi.org/10.1103/PhysRevLett.120.231601}{\emph{Phys. Rev. Lett.}
  {\bfseries 120} (2018) 231601}
  [\href{https://arxiv.org/abs/1710.00921}{{\ttfamily 1710.00921}}].

\bibitem{Grozdanov:2018kkt}
S.~Grozdanov, \emph{{On the connection between hydrodynamics and quantum chaos
  in holographic theories with stringy corrections}},
  \href{https://doi.org/10.1007/JHEP01(2019)048}{\emph{JHEP} {\bfseries 01}
  (2019) 048} [\href{https://arxiv.org/abs/1811.09641}{{\ttfamily
  1811.09641}}].

\bibitem{Guo:2019csw}
H.~Guo, Y.~Gu and S.~Sachdev, \emph{{Transport and chaos in lattice
  Sachdev-Ye-Kitaev models}},
  \href{https://doi.org/10.1103/PhysRevB.100.045140}{\emph{Phys. Rev. B}
  {\bfseries 100} (2019) 045140}
  [\href{https://arxiv.org/abs/1904.02174}{{\ttfamily 1904.02174}}].

\bibitem{Grozdanov:2019uhi}
S.~Grozdanov, P.K.~Kovtun, A.O.~Starinets and P.~Tadi\'c, \emph{{The complex
  life of hydrodynamic modes}},
  \href{https://doi.org/10.1007/JHEP11(2019)097}{\emph{JHEP} {\bfseries 11}
  (2019) 097} [\href{https://arxiv.org/abs/1904.12862}{{\ttfamily
  1904.12862}}].

\bibitem{Natsuume:2019xcy}
M.~Natsuume and T.~Okamura, \emph{{Nonuniqueness of Green's functions at
  special points}}, \href{https://doi.org/10.1007/JHEP12(2019)139}{\emph{JHEP}
  {\bfseries 12} (2019) 139}
  [\href{https://arxiv.org/abs/1905.12015}{{\ttfamily 1905.12015}}].

\bibitem{Natsuume:2019sfp}
M.~Natsuume and T.~Okamura, \emph{{Holographic chaos, pole-skipping, and
  regularity}}, \href{https://doi.org/10.1093/ptep/ptz155}{\emph{PTEP}
  {\bfseries 2020} (2020) 013B07}
  [\href{https://arxiv.org/abs/1905.12014}{{\ttfamily 1905.12014}}].

\bibitem{Ahn:2019rnq}
Y.~Ahn, V.~Jahnke, H.-S.~Jeong and K.-Y.~Kim, \emph{{Scrambling in Hyperbolic
  Black Holes: shock waves and pole-skipping}},
  \href{https://doi.org/10.1007/JHEP10(2019)257}{\emph{JHEP} {\bfseries 10}
  (2019) 257} [\href{https://arxiv.org/abs/1907.08030}{{\ttfamily
  1907.08030}}].

\bibitem{Natsuume:2019vcv}
M.~Natsuume and T.~Okamura, \emph{{Pole-skipping with finite-coupling
  corrections}}, \href{https://doi.org/10.1103/PhysRevD.100.126012}{\emph{Phys.
  Rev. D} {\bfseries 100} (2019) 126012}
  [\href{https://arxiv.org/abs/1909.09168}{{\ttfamily 1909.09168}}].

\bibitem{Wu:2019esr}
X.~Wu, \emph{{Higher curvature corrections to pole-skipping}},
  \href{https://doi.org/10.1007/JHEP12(2019)140}{\emph{JHEP} {\bfseries 12}
  (2019) 140} [\href{https://arxiv.org/abs/1909.10223}{{\ttfamily
  1909.10223}}].

\bibitem{Ceplak:2019ymw}
N.~Ceplak, K.~Ramdial and D.~Vegh, \emph{{Fermionic pole-skipping in
  holography}}, \href{https://doi.org/10.1007/JHEP07(2020)203}{\emph{JHEP}
  {\bfseries 07} (2020) 203}
  [\href{https://arxiv.org/abs/1910.02975}{{\ttfamily 1910.02975}}].

\bibitem{Abbasi:2019rhy}
N.~Abbasi and J.~Tabatabaei, \emph{{Quantum chaos, pole-skipping and
  hydrodynamics in a holographic system with chiral anomaly}},
  \href{https://doi.org/10.1007/JHEP03(2020)050}{\emph{JHEP} {\bfseries 03}
  (2020) 050} [\href{https://arxiv.org/abs/1910.13696}{{\ttfamily
  1910.13696}}].

\bibitem{Liu:2020yaf}
Y.~Liu and A.~Raju, \emph{{Quantum Chaos in Topologically Massive Gravity}},
  \href{https://arxiv.org/abs/2005.08508}{{\ttfamily 2005.08508}}.

\bibitem{Ahn:2020bks}
Y.~Ahn, V.~Jahnke, H.-S.~Jeong, K.-Y.~Kim, K.-S.~Lee and M.~Nishida,
  \emph{{Pole-skipping of scalar and vector fields in hyperbolic space:
  conformal blocks and holography}},
  \href{https://arxiv.org/abs/2006.00974}{{\ttfamily 2006.00974}}.

\bibitem{Abbasi:2020ykq}
N.~Abbasi and S.~Tahery, \emph{{Complexified quasinormal modes and the
  pole-skipping in a holographic system at finite chemical potential}},
  \href{https://arxiv.org/abs/2007.10024}{{\ttfamily 2007.10024}}.

\bibitem{Jansen:2020hfd}
A.~Jansen and C.~Pantelidou, \emph{{Quasinormal modes in charged fluids at
  complex momentum}},  \href{https://arxiv.org/abs/2007.14418}{{\ttfamily
  2007.14418}}.

\bibitem{Grozdanov:2020koi}
S.~Grozdanov, \emph{{Bounds on transport from univalence and pole-skipping}},
  \href{https://arxiv.org/abs/2008.00888}{{\ttfamily 2008.00888}}.

\bibitem{Gu:2016oyy}
Y.~Gu, X.-L.~Qi and D.~Stanford, \emph{{Local criticality, diffusion and chaos
  in generalized Sachdev-Ye-Kitaev models}},
  \href{https://doi.org/10.1007/JHEP05(2017)125}{\emph{JHEP} {\bfseries 05}
  (2017) 125} [\href{https://arxiv.org/abs/1609.07832}{{\ttfamily
  1609.07832}}].

\bibitem{Gu:2017ohj}
Y.~Gu, A.~Lucas and X.-L.~Qi, \emph{{Energy diffusion and the butterfly effect
  in inhomogeneous Sachdev-Ye-Kitaev chains}},
  \href{https://doi.org/10.21468/SciPostPhys.2.3.018}{\emph{SciPost Phys.}
  {\bfseries 2} (2017) 018} [\href{https://arxiv.org/abs/1702.08462}{{\ttfamily
  1702.08462}}].

\bibitem{Gu:2017njx}
Y.~Gu, A.~Lucas and X.-L.~Qi, \emph{{Spread of entanglement in a
  Sachdev-Ye-Kitaev chain}},
  \href{https://doi.org/10.1007/JHEP09(2017)120}{\emph{JHEP} {\bfseries 09}
  (2017) 120} [\href{https://arxiv.org/abs/1708.00871}{{\ttfamily
  1708.00871}}].

\bibitem{Haehl:2018izb}
F.M.~Haehl and M.~Rozali, \emph{{Effective Field Theory for Chaotic CFTs}},
  \href{https://doi.org/10.1007/JHEP10(2018)118}{\emph{JHEP} {\bfseries 10}
  (2018) 118} [\href{https://arxiv.org/abs/1808.02898}{{\ttfamily
  1808.02898}}].

\bibitem{Belin:2017jli}
A.~Belin, \emph{{Permutation Orbifolds and Chaos}},
  \href{https://doi.org/10.1007/JHEP11(2017)131}{\emph{JHEP} {\bfseries 11}
  (2017) 131} [\href{https://arxiv.org/abs/1705.08451}{{\ttfamily
  1705.08451}}].

\bibitem{Son:2002sd}
D.T.~Son and A.O.~Starinets, \emph{{Minkowski space correlators in AdS / CFT
  correspondence: Recipe and applications}},
  \href{https://doi.org/10.1088/1126-6708/2002/09/042}{\emph{JHEP} {\bfseries
  09} (2002) 042} [\href{https://arxiv.org/abs/hep-th/0205051}{{\ttfamily
  hep-th/0205051}}].

\bibitem{Hawking:1982dh}
S.~Hawking and D.N.~Page, \emph{{Thermodynamics of Black Holes in anti-De
  Sitter Space}}, \href{https://doi.org/10.1007/BF01208266}{\emph{Commun. Math.
  Phys.} {\bfseries 87} (1983) 577}.

\bibitem{Brown:1986nw}
J.~Brown and M.~Henneaux, \emph{{Central Charges in the Canonical Realization
  of Asymptotic Symmetries: An Example from Three-Dimensional Gravity}},
  \href{https://doi.org/10.1007/BF01211590}{\emph{Commun. Math. Phys.}
  {\bfseries 104} (1986) 207}.

\bibitem{Keller:2011xi}
C.A.~Keller, \emph{{Phase transitions in symmetric orbifold CFTs and
  universality}}, \href{https://doi.org/10.1007/JHEP03(2011)114}{\emph{JHEP}
  {\bfseries 03} (2011) 114} [\href{https://arxiv.org/abs/1101.4937}{{\ttfamily
  1101.4937}}].

\bibitem{Belin:2014fna}
A.~Belin, C.A.~Keller and A.~Maloney, \emph{{String Universality for
  Permutation Orbifolds}},
  \href{https://doi.org/10.1103/PhysRevD.91.106005}{\emph{Phys. Rev. D}
  {\bfseries 91} (2015) 106005}
  [\href{https://arxiv.org/abs/1412.7159}{{\ttfamily 1412.7159}}].

\bibitem{Haehl:2014yla}
F.M.~Haehl and M.~Rangamani, \emph{{Permutation orbifolds and holography}},
  \href{https://doi.org/10.1007/JHEP03(2015)163}{\emph{JHEP} {\bfseries 03}
  (2015) 163} [\href{https://arxiv.org/abs/1412.2759}{{\ttfamily 1412.2759}}].

\bibitem{Belin:2015hwa}
A.~Belin, C.A.~Keller and A.~Maloney, \emph{{Permutation Orbifolds in the large
  N Limit}}, \href{https://doi.org/10.1007/s00023-016-0529-y}{\emph{Annales
  Henri Poincare} (2016) 1} [\href{https://arxiv.org/abs/1509.01256}{{\ttfamily
  1509.01256}}].

\bibitem{Hartman:2015lfa}
T.~Hartman, S.~Jain and S.~Kundu, \emph{{Causality Constraints in Conformal
  Field Theory}}, \href{https://doi.org/10.1007/JHEP05(2016)099}{\emph{JHEP}
  {\bfseries 05} (2016) 099}
  [\href{https://arxiv.org/abs/1509.00014}{{\ttfamily 1509.00014}}].

\bibitem{streater2000pct}
R.F.~Streater and A.S.~Wightman, \emph{PCT, spin and statistics, and all that},
  vol.~52, Princeton University Press (2000).

\bibitem{haag2012local}
R.~Haag, \emph{Local quantum physics: Fields, particles, algebras}, Springer
  Science \& Business Media (2012).

\bibitem{Eguchi:1986sb}
T.~Eguchi and H.~Ooguri, \emph{{Conformal and Current Algebras on General
  Riemann Surface}},
  \href{https://doi.org/10.1016/0550-3213(87)90686-9}{\emph{Nucl. Phys. B}
  {\bfseries 282} (1987) 308}.

\bibitem{Halder:2019ric}
I.~Halder, \emph{{Global Symmetry and Maximal Chaos}},
  \href{https://arxiv.org/abs/1908.05281}{{\ttfamily 1908.05281}}.

\bibitem{Birmingham:2001pj}
D.~Birmingham, I.~Sachs and S.N.~Solodukhin, \emph{{Conformal field theory
  interpretation of black hole quasinormal modes}},
  \href{https://doi.org/10.1103/PhysRevLett.88.151301}{\emph{Phys. Rev. Lett.}
  {\bfseries 88} (2002) 151301}
  [\href{https://arxiv.org/abs/hep-th/0112055}{{\ttfamily hep-th/0112055}}].

\bibitem{Cho:2017fzo}
M.~Cho, S.~Collier and X.~Yin, \emph{{Genus Two Modular Bootstrap}},
  \href{https://doi.org/10.1007/JHEP04(2019)022}{\emph{JHEP} {\bfseries 04}
  (2019) 022} [\href{https://arxiv.org/abs/1705.05865}{{\ttfamily
  1705.05865}}].

\bibitem{Haehl:2019eae}
F.M.~Haehl, W.~Reeves and M.~Rozali, \emph{{Reparametrization modes, shadow
  operators, and quantum chaos in higher-dimensional CFTs}},
  \href{https://doi.org/10.1007/JHEP11(2019)102}{\emph{JHEP} {\bfseries 11}
  (2019) 102} [\href{https://arxiv.org/abs/1909.05847}{{\ttfamily
  1909.05847}}].

\end{thebibliography}\endgroup
\bibliographystyle{JHEP}

\end{document}